\documentclass[useAMS,usenatbib]{mn2e}
\usepackage{times}
\usepackage{graphics,epsfig}
\usepackage{graphicx}
\usepackage{amssymb}
\title[Constraining MOND from spherical accretion]{Constraining the MOdified Newtonian Dynamics from spherically symmetrical hydrodynamic accretion} 
\author[N. Roy]{Nirupam Roy\thanks{E-mail: nroy@aoc.nrao.edu}\\ 
National Radio Astronomy Observatory, P. O. Box O, 1003 Lopezville Road, Socorro, NM 87801, USA}

\begin{document}
\date{Accepted yyyy month dd. Received yyyy month dd; in original form yyyy 
month dd}

\pagerange{\pageref{firstpage}--\pageref{lastpage}} \pubyear{2010}

\maketitle

\label{firstpage}

\begin{abstract}
The MOdified Newtonian Dynamics (MOND) is an alternative to the dark matter 
assumption that can explain the observed flat rotation curve of galaxies. 
Here hydrodynamic accretion is considered to critically check the consistency 
and to constrain the physical interpretation of this theory. It is found that, 
in case of spherically symmetrical hydrodynamic accretion, the modified 
Euler's equation has real solution if the interpretation is assumed to be a 
modification of the law of dynamics. There is no real solution if it is 
assumed to be an acceleration scale dependent modification of the 
gravitational law. With the modified Euler's equation, the steady state mass 
accretion rate is found to change up to $\sim 15\%$. The astrophysical and 
cosmological implications of these results are also discussed.
\end{abstract}


\section{Introduction}
\label{sec:int}

The rotation curve of spiral galaxies can not be explained in terms of 
gravitational potential of only the visible mass with any reasonable mass 
to light ratio \citep[e.g.][]{rubin70,rubin80,sofue96,sofue01,spano08}. 
This indicates the presence of a significant fraction of gravitating 
matter with very high mass to light ratio. Various other observational 
and theoretical constraints from velocity dispersion of elliptical 
galaxies, baryon fraction in galaxy clusters, gravitational lensing, 
structure formation, CMB power spectrum, observation of Lyman $\alpha$ 
forest etc. \citep[e.g.][]{faber76,wu98,spring05,clowe06,mass07,viel09} 
also point out that a significant fraction of the mass of the Universe is 
in the form of the dark matter with very little or no electromagnetic 
interaction \citep{bertone05}. Though the dark matter concept is widely 
accepted to be an explanation of these observations, there is no general 
agreement on the composition and various other properties of this major 
constituent of the Universe \citep[e.g.][]{vs84,defw85,ui85,nfw96,blok05}. 
Though there have been indications from some of the direct and indirect 
detection experiments of different types of dark matters, no conclusive 
results showing firm detection is still reported.

However, over the time, a variety of alternative theories, with 
some modification of either laws of motion or law of gravitational 
force, have been suggested to explain some of the observations without 
invoking the assumption of any dark matter 
\citep[e.g.][]{milgrom83a,beken84,sand86,fahr90,sanders97,brown06}. A 
particularly successful theory in this category is the theory of the 
MOdified Newtonian Dynamics (MOND) proposed to explain rotation curves 
without and ``hidden mass hypothesis'' \citep{milgrom83a,milgrom83b}. 
Essentially, the proposal of this theory is that the acceleration due to 
gravitational force is not linearly proportional to the force at very low 
acceleration limit. This simple modification has so far successfully 
explained most of the observed rotation curves for different types of 
galaxies \citep[e.g.][]{sr03,ssm10}, and the few cases where it fails can 
be explained in terms of inadequate data, large asymmetries in the velocity 
field  and observational uncertainties \citep{milgrom91,sw99}. The MOND 
predictions have been independently tested against observations 
\citep[e.g.][and references therein]{fbz07,tc07,sm02} and found to be 
consistent. Though there have been some criticism of this theory followed 
by the claim of a direct proof of dark matter at cluster scale from weak 
lensing observations \citep{clowe06}, there are ways of accommodating these 
observations \citep[e.g.][]{angus06,angus07}, and the issue is far from 
being settled.

Whether or not MOND is an alternative to the dark matter scenario, there 
is no doubt that a complete theory of dark matter must explain this success 
of MOND in predicting and explaining such observations. It is, hence, 
increasingly important to critically verify its consistency at all scales. 
There are proposals to test predictions of MOND or associated theories in a 
variety of ways \citep[e.g.][]{delo09,trenkel10}. In this paper, I have 
considered the astrophysical case of spherical accretion to check the 
consistency and to constrain the physical interpretation of MOND. The 
background is outlined in Section \S\ref{sec:bkg}, and  the outcome in MOND 
regime for different modifications are described in Section \S\ref{sec:mda}. 
Section \S\ref{sec:dis} contains discussions on the results, and I present 
conclusions in Section \S\ref{sec:con}.

\section{Background}
\label{sec:bkg}

In MOND, the Newtonian equation of dynamics $\vec{F} = m\vec{a}$ is modified 
to $\vec{F} = m\mu\vec{a}$, where $\mu = \mu(a)$ is a dimensionless parameter. 
This modification is significant at very low acceleration regime (below an 
acceleration of $a_0 \approx 10^{-10}$ m~s$^{-1}$). It is proposed that $\mu 
\approx a/a_0$ for $a < a_0$ and $\mu = 1$ for $a > a_0$, and the exact form 
of $\mu(a)$ may not have any serious consequences. Thus the motion due to 
gravitational force will be governed by
\begin{equation}
\vec{F}_g = \frac{GMm}{r^2}\hat{r} = m\mu \vec{a}.
\label{eqn:mond}
\end{equation}
For the low acceleration at a large distance from the centre of a galaxy, 
this will imply $a = v^2/r = \sqrt{GMa_0}/r$ giving rise to a flat rotation 
curve. See \citet{sm02} for a comprehensive critical review of the theory, a 
more generalized formulation and its implications, its observational supports 
and other details. Also see \citet{beke04} and references therein for the 
details of the ``MOND inspired'' relativistic, generalized theory of 
gravitation called TeVeS. For the purpose of this work, I will only use this 
theory to be phenomenological as summarized in equation (\ref{eqn:mond}) to 
investigate its possible implications in the case of spherically symmetrical 
hydrodynamic accretion.

Note that in equation (\ref{eqn:mond}), $\mu$ can be written as a modification 
of either the inertial term ($\vec{F}_g = m\mu \vec{a}$) or modification of 
the gravitation force term ($\vec{F}_g/\mu = m\vec{a}$). Though they lead to 
the same result for the rotation curve, the physical implication is 
significantly different. There may be situations where these two 
interpretations leads to drastically different results. Note that this 
phenomenological description of the dynamics is consistent with the 
nonrelativistic limit of TeVeS with spherical symmetry \citep{beke04}.

\section{Accretion in MOND regime}
\label{sec:mda}

Here I have considered spherical accretion to check if, in the MOND regime, 
there is any change of physical conditions from that of the Newtonian 
scenario. In the Newtonian case, the governing equations for spherically 
symmetrical hydrodynamic steady state accretion are the continuity equation 
and the Euler's equation
\begin{equation}
\frac{1}{r^2}\frac{d}{dr}(\rho ur^2) = 0
\label{ncont}
\end{equation}
\begin{equation}
u\frac{du}{dr} = -\frac{1}{\rho}\frac{dP}{dr}-\frac{GM}{r^2}
\label{neuler}
\end{equation}
where $u(r)$ is the radial inward velocity, $P(r)$ and $\rho(r)$ are pressure 
and density related by an equation of state $P = K\rho^\gamma$ and polytropic 
index $\gamma$, and $M$ is the mass of accretor. The sound speed in the medium 
$c_s(r)$ is related to $P$ and $\rho$ as $c_s^2 = dP/d\rho$.

Starting with a boundary condition $\rho_{\infty}$ and $c_{s\infty}$ at a very 
large distance from the central accretor, equations (\ref{ncont}) and 
(\ref{neuler}) can be solved for a given $M$ and $\gamma$ to derive the steady 
state density and velocity profile. The solution of astrophysical interest is 
an unique transonic solution with a mass accretion rate of $\dot{M} = 
4\pi\lambda(\gamma)\rho_{\infty}G^2 M^2/c^3_{s\infty}$, where $\lambda$ is a 
dimensionless constant. Note that the acceleration at the sonic point is of 
the order of $c^4_{s\infty}/GM$. For a solar mass accretor and a typical 
ambient sound speed of $\sim 10$ km~s$^{-1}$, this is $\sim 10^{-7}$, very 
much comparable to the MOND acceleration constant $a_0$. Keeping this in mind, 
it is useful here to introduce a dimensionless parameter $\widetilde{a}_0 = 
a_0/(c^4_{s\infty}/GM)$ and to study the behaviour of the system for different 
values of this parameter.

In case of the hydrodynamic accretion, the dynamics is governed by the 
interplay of three terms - pressure, gravitational force and inertia. In the 
aforementioned two different interpretations of the MOND modification, viz. 
modification of gravitational force and modification of the Newtonian dynamics, 
these three terms change in different ways. Thus, considering this case of 
hydrodynamic accretion gives us a chance to study any possible difference that 
may arise in these two interpretations. 

\subsection{Modification of gravitational force}

\begin{figure}
\begin{center}
\includegraphics[scale=0.65, angle=0.0]{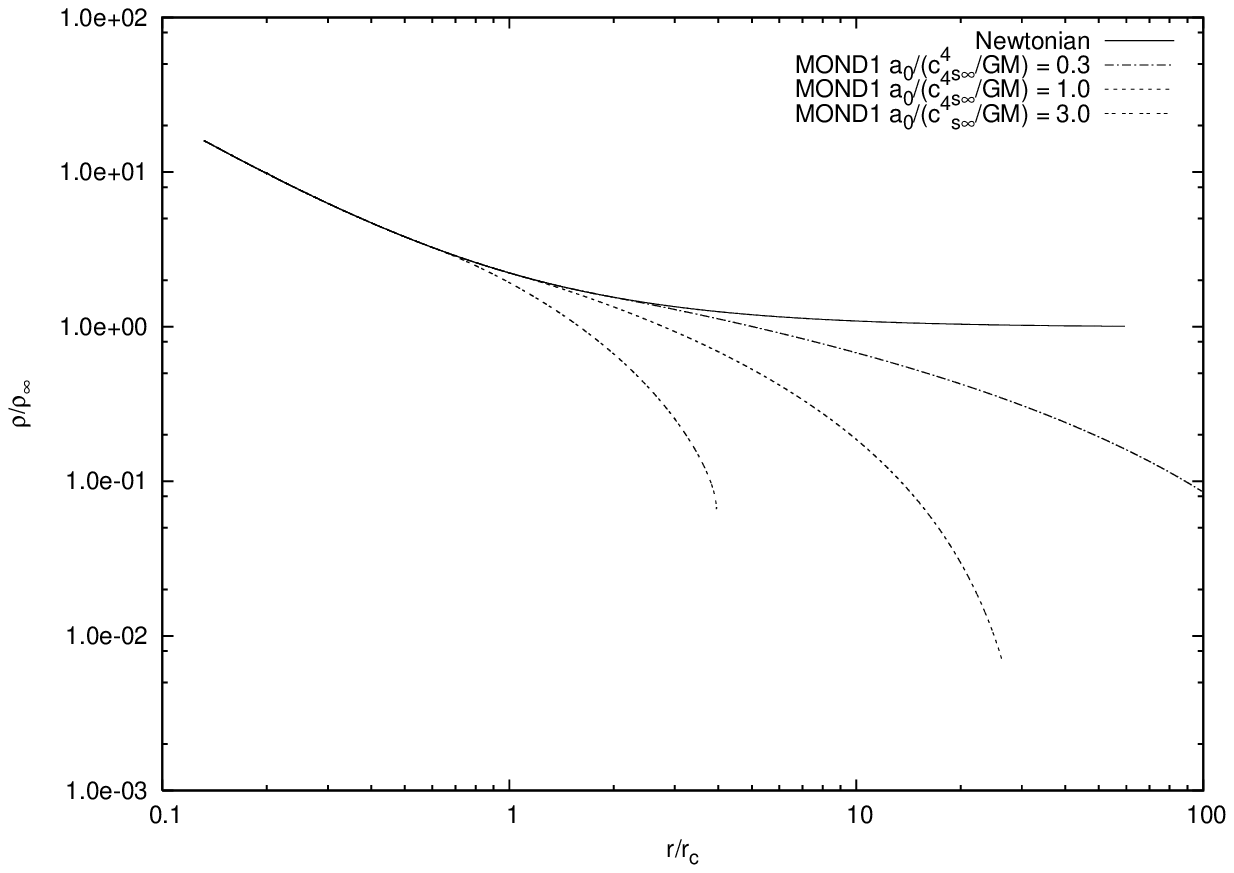}
\includegraphics[scale=0.65, angle=0.0]{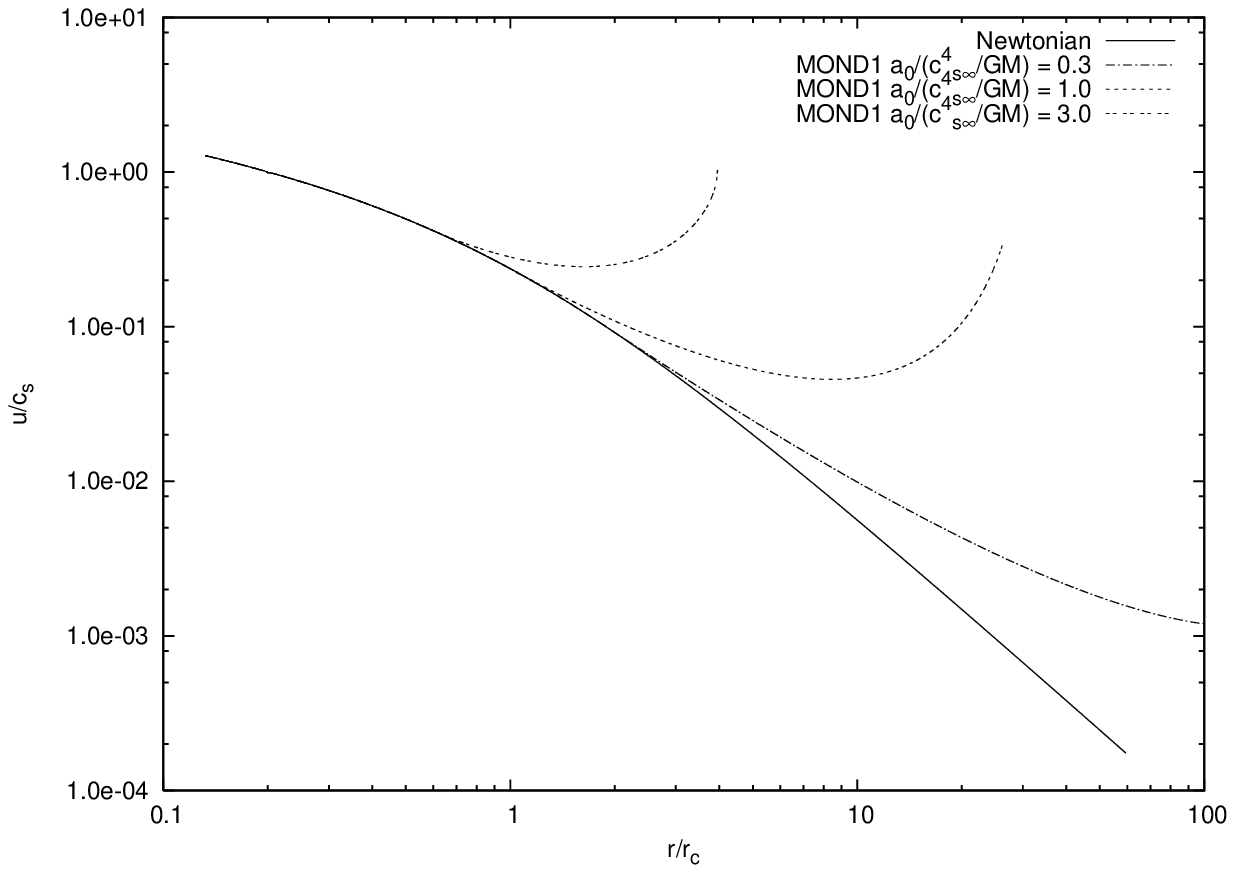}
\caption{\label{fig:mond-1} Density and velocity profile in the MOND regime for the modified gravitational force case. Here $\gamma = 7/5$ and $\widetilde{a}_0 \approx 0.3$, $1.0$ and $3.0$ (for $c_{s\infty}$ of the Newtonian solution). Solid line is for the Newtonian solution.}
\end{center}
\end{figure}

In this case, we consider the modification of the gravitational force of the 
form $\vec{F}^\prime_g = \vec{F}_g/\mu = m\vec{a}$. As mentioned earlier, this 
can be derived from TeVeS at nonrelativistic limit assuming spherical 
symmetry \citep{beke04}. With this modification, the continuity equation will 
remain unchanged from equation (\ref{ncont}), but the Euler's equation will be 
modified to 
\begin{equation}
u\frac{du}{dr} = -\frac{1}{\rho}\frac{dP}{dr}-\frac{GM}{\mu r^2}
\label{m1euler}
\end{equation}
where $\mu = \mu(a/a_0)$ and $a = \sqrt{GMa_0}/r$. Note that, in MOND regime, 
with $\mu(a/a_0) = a/a_0 < 1$, the gravitational term will be $\sqrt{GMa_0}/r$, 
whereas in the low acceleration regime with $mu = 1$, it will be same as the 
regular Newtonian term $GM/r^2$. Using equation (\ref{ncont}) and the equation 
of state, $P$ and $\rho$ can be eliminated from equation (\ref{m1euler}) in 
terms of $u$, and can be re-written as
\begin{equation}
\frac{(u^2-c^2_s)}{u}\frac{du}{dr} = \frac{2c^2_s}{r} - \frac{GM}{\mu r^2}.
\label{dudr1}
\end{equation}
With an initial condition $\rho_i$ and $a_i$ at a small radius $r_i$, equation 
(\ref{dudr1}), along with the equation of state and the continuity equation, 
can be solved numerically for the density and velocity profile. Note that for 
typical astrophysical condition, $a$ at small radius is significantly larger 
than $a_0$. So, for the inner region, the MONDian solution will not differ 
from the Newtonian solution. Here also, like the Newtonian case, velocity at 
$r_i$ should have a unique value for the solution to pass through the sonic 
point and to give the right accretion rate. For a given $\gamma$ and 
$\widetilde{a}_0$, equation (\ref{dudr1}) can be solved with same initial 
conditions for both the Newtonian and the MONDian case. For the Newtonian 
case, at large radius, $\rho$ and $c_s$ tends asymptotically to a constant 
value $\rho_{\infty}$ and $c_{s\infty}$ respectively, and the velocity $u 
\sim r^{-2}$ tends to zero. Interestingly, as shown in figure 
(\ref{fig:mond-1}), for MONDian case, equation (\ref{dudr1}) does not have 
this asymptotic solution at large radius. Figure (\ref{fig:mond-1}) shows the 
Newtonian and MONDian solutions for $\gamma = 7/5$ and $\widetilde{a}_0 
\approx 0.3$, $1.0$ and $3.0$. The top and bottom panels show the density and 
the velocity field respectively. Both the density and the velocity profile 
deviate from the Newtonian solution and diverges away from the asymptotic 
solution at large radius. In these plots, the density and the velocity values 
are scaled by $\rho_\infty$ and $c_{s\infty}$ (of the Newtonian solution), and 
the radius is scaled by $r_c = GM/c^2{s\infty}$.

One can understand this result analytically with the following arguments. In 
the MOND regime with $\mu = a/a_0 < 1$, equation (\ref{dudr1}) can be written 
in terms of $u^\prime = \frac{du}{dr}$ as 
\begin{equation}
{u^\prime} = u\frac{(\frac{2c^2_s}{r} - \frac{\sqrt{GMa_0}}{r})}{(u^2-c^2_s)}.
\label{dudr2}
\end{equation}
Since both the term in numerator scales as $1/r$, with decreasing $c_s$, 
$u^\prime$ changes sign at large radius. Thus, $u/c_s$ is no more a 
monotonically decreasing function with increasing radius, and the MONDian 
solution diverges from the Newtonian solution. It implies that there is no 
solution of the flow with $\rho$, $u$ and $c_s$ having the right physical 
boundary condition.

\subsection{Modification of dynamics}

\begin{figure}
\begin{center}
\includegraphics[scale=0.65, angle=0.0]{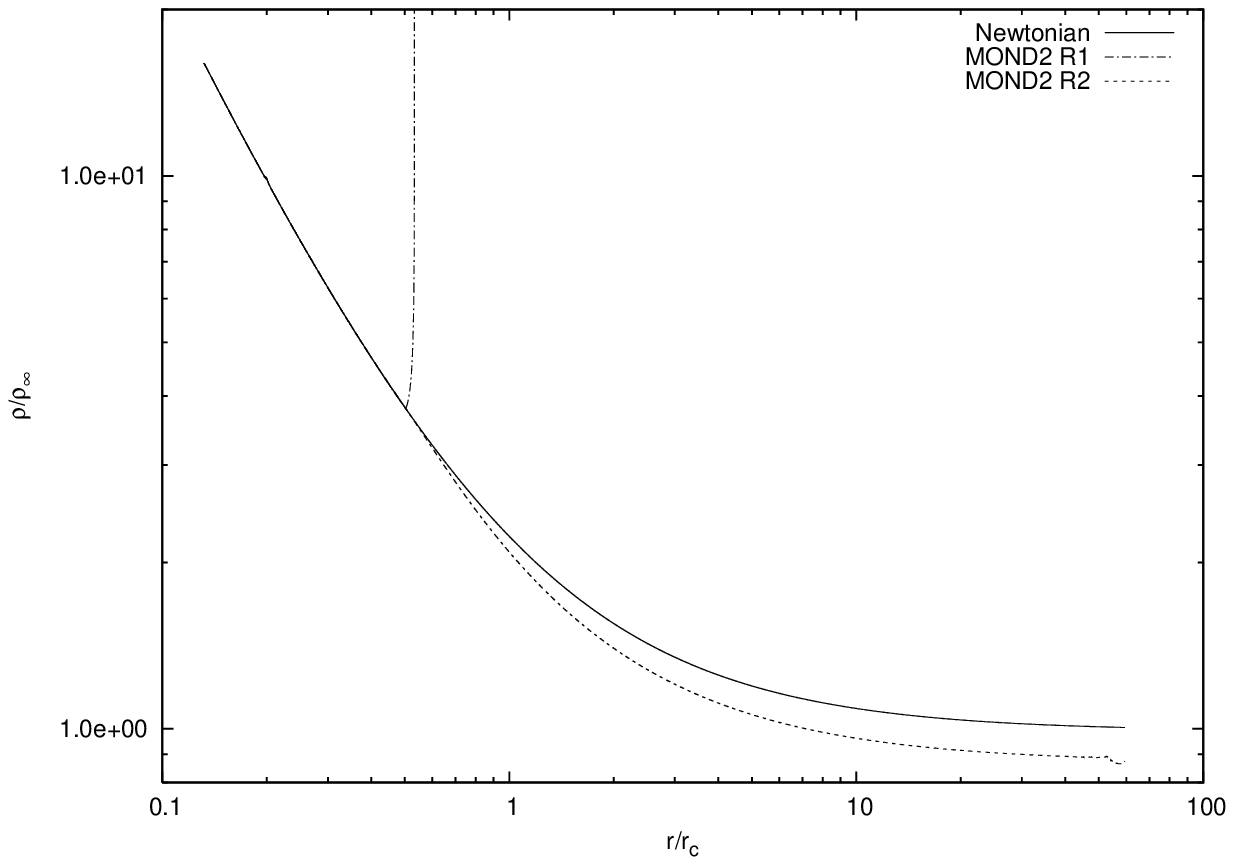}
\includegraphics[scale=0.65, angle=0.0]{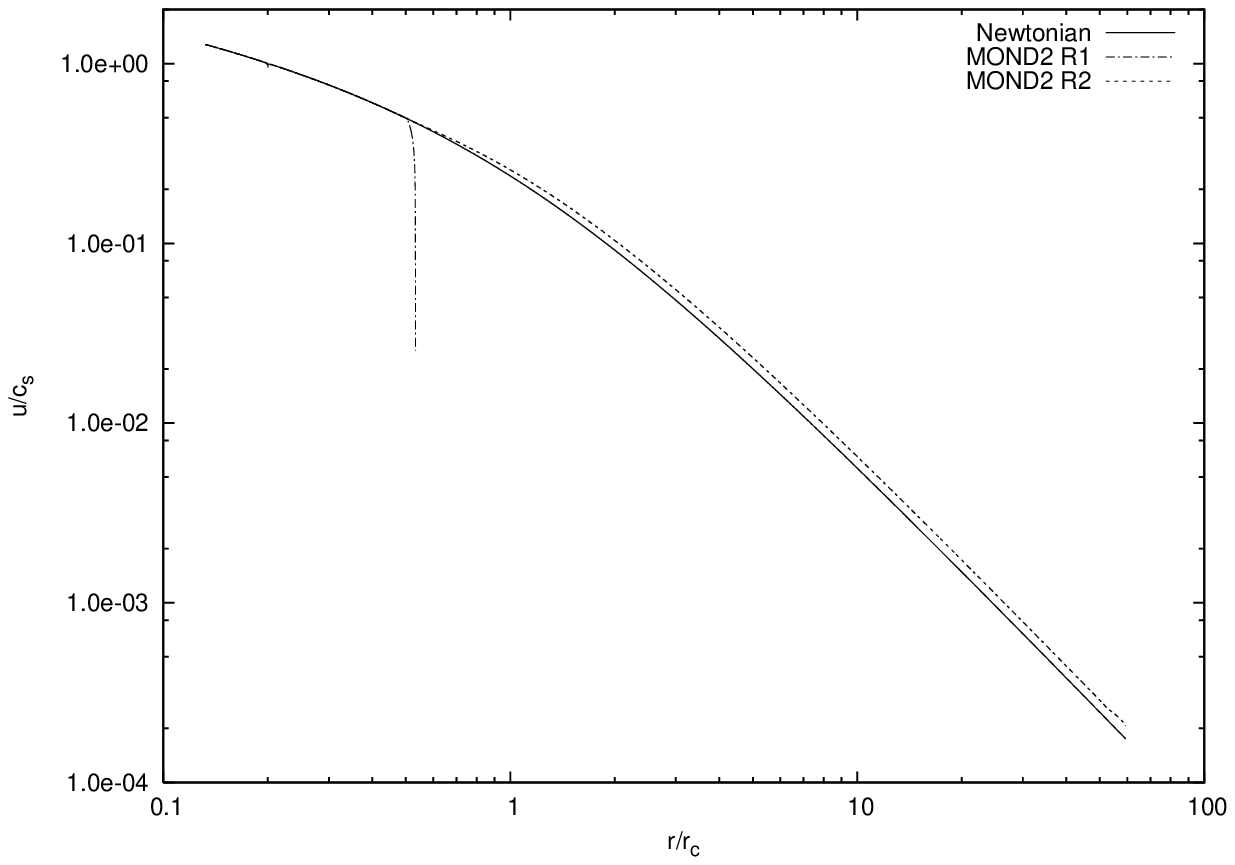}
\caption{\label{fig:mond-2} Typical spherical accretion density and velocity profile in the MOND regime for the modified dynamics case. Here $\gamma = 7/5$ and $\widetilde{a}_0 \approx 1$ (for $c_{s\infty}$ of the Newtonian solution). Solid line is for the Newtonian solution. Other two broken lines correspond to the two different roots of equation (\ref{dudr4}). One of the roots is imaginary at large radius but the other one satisfies physical boundary condition requirement.}
\end{center}
\end{figure}

In this case also, the continuity equation will remain as in equation 
(\ref{ncont}). But since $F = m\mu a$, the modified Euler's equation will be
\begin{equation}
u\mu\frac{du}{dr} = -\frac{1}{\rho}\frac{dP}{dr}-\frac{GM}{r^2}.
\label{m2euler}
\end{equation}
Following a similar numerical analysis as in the earlier case, one can solve 
for $u(r)$ and $\rho(r)$ from equations (\ref{ncont}) and (\ref{m2euler}). As 
shown in figure (\ref{fig:mond-2}), one of the solutions in this case starts 
to deviate from that of the Newtonian one at a large radius, and have 
asymptotic value of $\rho_\infty$ and $c_{s\infty}$ lower than that of the 
Newtonian value with same mass accretion rate. The other solution becomes 
imaginary after a certain distance. Figure (\ref{fig:mond-2}) shows the 
Newtonian and these two MONDian solution for $\gamma = 7/5$ and 
$\widetilde{a}_0 \approx 1$. The deviation of $\rho_\infty$ from that of the 
Newtonian one depends on the value of both $\gamma$ and $\widetilde{a}_0$.
 
Eliminating $P$ and $\rho$ using equation (\ref{ncont}) and the equation of 
state, equation (\ref{m2euler}) can be rewritten as
\begin{equation}
\frac{\mu u^2-c^2_s}{u}\frac{du}{dr} = \frac{2c^2_s}{r}-\frac{GM}{r^2}.
\label{dudr3}
\end{equation}
In the MOND regime, where $\mu = a/a_0 = -\frac{u}{a_0}\frac{du}{dr}$, this 
also becomes a quadratic equation in $u^\prime$
\begin{equation}
\frac{u^2}{a_0}{u^\prime}^2 +\frac{c^2_s}{u}u^\prime + (\frac{2c^2_s}{r} - \frac{GM}{r^2})=0
\label{dudr4}
\end{equation}
where the condition of a real solution for $u^\prime$ is
\begin{equation}
\frac{u^4}{c^4_s} \leq \frac{a_0}{4}(\frac{2c^2_s}{r} - \frac{GM}{r^2})^{-1}.
\label{cond2}
\end{equation}
This is an upper limit condition to $u$ and does not contradict the required 
physical condition of velocity tending to zero at large radius. Thus, in this 
case, there exist a physically meaningful solution where $u$ tends to zero at 
large radius whereas density and sound speed asymptotically tend to 
$\rho_\infty$ and $c_{s\infty}$ respectively.

\begin{figure}
\begin{center}
%GNUPLOT: LaTeX picture with Postscript
\begin{picture}(0,0)%
\includegraphics{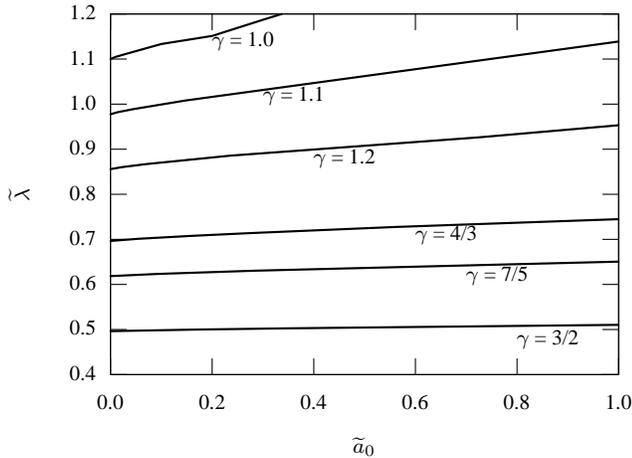}%
\end{picture}%
\begingroup
\setlength{\unitlength}{0.0200bp}%
\begin{picture}(12600,9000)(0,0)%
\put(1925,1650){\makebox(0,0)[r]{\strut{}0.4}}%
\put(1925,2500){\makebox(0,0)[r]{\strut{}0.5}}%
\put(1925,3350){\makebox(0,0)[r]{\strut{}0.6}}%
\put(1925,4200){\makebox(0,0)[r]{\strut{}0.7}}%
\put(1925,5050){\makebox(0,0)[r]{\strut{}0.8}}%
\put(1925,5900){\makebox(0,0)[r]{\strut{}0.9}}%
\put(1925,6750){\makebox(0,0)[r]{\strut{}1.0}}%
\put(1925,7600){\makebox(0,0)[r]{\strut{}1.1}}%
\put(1925,8450){\makebox(0,0)[r]{\strut{}1.2}}%
\put(2200,1100){\makebox(0,0){\strut{}0.0}}%
\put(4115,1100){\makebox(0,0){\strut{}0.2}}%
\put(6030,1100){\makebox(0,0){\strut{}0.4}}%
\put(7945,1100){\makebox(0,0){\strut{}0.6}}%
\put(9860,1100){\makebox(0,0){\strut{}0.8}}%
\put(11775,1100){\makebox(0,0){\strut{}1.0}}%
\put(550,5050){\rotatebox{90}{\makebox(0,0){\strut{}$\widetilde{\lambda}$}}}%
\put(6987,275){\makebox(0,0){\strut{}$\widetilde{a}_0$}}%
\put(4115,7940){\makebox(0,0)[l]{\strut{}$\gamma$ = 1.0}}%
\put(5073,6920){\makebox(0,0)[l]{\strut{}$\gamma$ = 1.1}}%
\put(6030,5730){\makebox(0,0)[l]{\strut{}$\gamma$ = 1.2}}%
\put(7945,4285){\makebox(0,0)[l]{\strut{}$\gamma$ = 4/3}}%
\put(8903,3520){\makebox(0,0)[l]{\strut{}$\gamma$ = 7/5}}%
\put(9860,2330){\makebox(0,0)[l]{\strut{}$\gamma$ = 3/2}}%
\end{picture}%
\endgroup
%\endinput
%\input{fig5.tex}
\caption{\label{fig:mond-3} Mass accretion rate parameter $\widetilde{\lambda}$ as a function of acceleration parameter $\widetilde{a}_0$ for different equation of state with polytropic index $\gamma$.}
\end{center}
\end{figure}

Effectively, for this interpretation, the general nature of the solution in 
MOND regime does not change from that of the Newtonian solution. In details, 
however, the mass accretion rate changes to $\dot{M} = 4\pi\widetilde{\lambda} 
(\gamma,a_0)\rho_{\infty}G^2 M^2/c_{s\infty}$, where $\widetilde{\lambda}$ is 
a dimensionless factor. This is further investigated by evaluating 
$\widetilde{\lambda}$ for different $\gamma$ and $\widetilde{a}_0$. As shown 
in figure (\ref{fig:mond-3}), the steady state mass accretion rate may change 
by up to $\sim 15\%$. However, in case of astrophysical accretion, the 
accretor mass, ambient density and sound speed are often not so well 
determined to observationally distinguish this change between Newtonian and 
MONDian accretion.

\section{Discussions}
\label{sec:dis}

The results derived in Section \S\ref{sec:mda} is for the spherically 
symmetrical and non-magnetized hydrodynamics accretion. While considering 
these results, it should be kept in mind that, in reality, astrophysical 
accretion is a complex phenomenon. These assumptions of steady state, 
spherical symmetry and no importance of angular momentum, and ignoring the 
possible role of self-gravity and magnetic field are for the simplicity of 
this semi-analytical investigation. However, in the low acceleration regime 
far away from the accretor, the effect of magnetic field, self gravity and 
angular momentum may be negligible, and the large scale accretion may be 
approximated as hydrodynamic, non-magnetized, spherical accretion onto a 
central compact accretor. Thus, the exact solution may get modified due to 
these complications, but the general nature of the solution is expected to 
remain unchanged.

The other point to note is that the pressure term in the Euler's equation is 
assumed to be not affected by the MONDian modification. Fundamentally, this 
term, which arises form the random motion and change of momentum of the 
particles, is expected to get modified in a similar way as the inertia term. 
In that case, equation (\ref{m2euler}) will be reduced to equation 
(\ref{m1euler}), and, there will be no real solution in the MOND regime. The 
way out is to consider equation (\ref{eqn:mond}) as the governing equation 
not for random motion but only for bulk motion of system with symmetry. This 
scenario is consistent with the observation that the random motion and 
acceleration of the gas does not alter the galaxy rotation curve either.

\section{Conclusions}
\label{sec:con}

For the case of spherically symmetrical hydrodynamic accretion in MOND regime, 
it is shown here that physically meaningful solution exists only for the 
interpretation of the modification of dynamics but not of the gravitational 
law. At a phenomenological level, this modification should be not for random 
motion but for bulk motion only. Given the uncertainty on various parameters, 
the change of accretion rate is not significant to distinguish between the 
Newtonian and the MONDian scenario.

\section*{Acknowledgements}

I am grateful to the anonymous referee for useful comments and for prompting 
us into improving this paper. I am also grateful to Sanjay Bhatnagar, Sayan 
Chakraborti, Susmita Chakravorty, Abhirup Datta and Prasun Dutta for for much 
encouragement. NR is a Jansky Fellow of the National Radio Astronomy 
Observatory (NRAO). The NRAO is a facility of the National Science Foundation 
operated under cooperative agreement by Associated Universities, Inc.

\bsp

\label{lastpage}

\end{document}